\documentclass[preprint, showpacs,preprintnumbers,amsmath,amssymb,nofootinbib]{revtex4}

 \usepackage{epsf}
 \usepackage{graphicx}

\begin{document}
\newcommand{\be}[1]{\begin{equation}\label{#1}}
 \newcommand{\ee}{\end{equation}}
 \newcommand{\bea}{\begin{eqnarray}}
 \newcommand{\eea}{\end{eqnarray}}
 \def\disp{\displaystyle}

 \def\gsim{ \lower .75ex \hbox{$\sim$} \llap{\raise .27ex \hbox{$>$}} }
 \def\lsim{ \lower .75ex \hbox{$\sim$} \llap{\raise .27ex \hbox{$<$}} }

\title{ Dynamics of interacting phantom scalar field dark energy in Loop Quantum Cosmology}

\author{Xiangyun Fu\;$^{1,2}$, Hongwei Yu\;$^{1,2,}$\footnote{Corresponding author} and Puxun Wu\;$^3$}
\affiliation{$^1$Department of Physics and Institute of  Physics,\\
Hunan Normal University, Changsha, Hunan 410081, China\\
$^2$ Key Laboratory of Low Dimensional Quantum Structures and
Quantum Control of Ministry of Education, Hunan Normal University ,
Changsha, Hunan 410081, P.R. China
 \\
$^3$Department of Physics and Tsinghua Center for Astrophysics,
Tsinghua University, Beijing 100084, China}

\begin{abstract}
We study the dynamics of a phantom scalar field dark energy
interacting with dark matter in loop quantum cosmology~(LQC). Two
kinds of coupling of the form $\alpha{\rho_m}{\dot\phi}$ (case I)
and $3\beta H (\rho_\phi +\rho_m)$ (case II) between the phantom
energy and dark matter are examined  with the potential for the
phantom field taken to be exponential. For both kinds of
interactions, we find that the future singularity appearing in the
standard FRW cosmology can be avoided by loop quantum gravity
effects.  In case II, if the
 phantom field is initially rolling down the potential,
the loop quantum effect has no influence on the cosmic late time
evolution and the universe will accelerate forever with a constant
energy ratio between the dark energy and dark matter.
\end{abstract}


\maketitle

\section{Introduction}\label{sec1}
Recently, present accelerating expansion of our universe has been
confirmed by many observations, such as the cosmic microwave
background (CMB) anisotropy, type Ia supernovae, and large scale
galaxy surveys~\cite{bennett,scranton}. In order to explain this
observed phenomena, dark energy is assumed to exist in the universe
within the framework of general relativity. Dark energy is an exotic
energy component with negative pressure and accounts for about
$72\%$ of present total cosmic energy.  In addition, observations
also show that there is another dark component in the universe,
i.e.,  dark matter, accounting for about $25\%$ of total cosmic
energy today. The simplest candidate of dark energy is the
cosmological constant. It is, however, plagued with the so-called
coincidence problem and the cosmological constant
problem~\cite{weinberg}. Thus some dynamical scalar fields, such as
quintessence~\cite{Quint}, phantom~\cite{Cald},
quintom~\cite{Quintom} and hessence~\cite{Hessence}, are proposed as
possible candidates of dark energy. It is worthy to note, however,
that for these scalar field models the coincidence problem still
remains. Although the two dark components are usually studied under
the assumption that there is no interaction between them, one can
not exclude such a possibility.  In fact, researches show that a
presumed interaction may help alleviate the coincidence
problem~\cite{Chimento2003}. Therefore interacting dark energy
models have attracted a great deal of interest.

 Most of the present studies on dark energy are carried out in the
framework of classical Einstein gravity. However, it is commonly
believed that quantum gravity effects would play a role in the
evolution of the universe. Therefore, it is desirable to examine the
properties of dark energy in a theory of quantum gravity. One of
such theories which we are interested in the present paper is Loop
Quantum Gravity (LQG) (see
e.g.~\cite{RovelliThiemann,Ashtekar2,Rovelli,Ashtekar3} for
reviews), which is a nonperturbative background independent theory.
At the quantum level, the classical spacetime continuum is replaced
by a discrete quantum geometry and the operators corresponding to
geometrical quantities have discrete eigenvalues.  LQG has been
applied in cosmological context as seen in various literature where
it is known as Loop Quantum Cosmology~(LQC)~\cite{r42,r43,
Mielczarek, r44}.  The effects of loop quantum gravity modify the
standard Friedmann equation by adding into it a correction term
$-\rho^{2}/\rho_{c}$ which essentially encodes the discrete quantum
geometric nature of spacetime~\cite{r44,r53, r54}. Let us note that
although extra terms such as those derived in Ref.~\cite{Bojowald08}
are in principle possible, the effect of such terms is negligible
for all practical purposes~\cite{SA}. When this correction term
becomes dominant, the universe begins to bounce and then oscillates
forever. Therefore both the future singularity and the singularity
at semi-classical regime can be avoided~\cite{r44, r53, Sami:2006wj,
Daris, Singh:2003au}. Recently, the dynamics of phantom, quintom and
hessence in loop quantum cosmology have been studied~\cite{Daris,
Hao,WuZhang} and behaviors different from that in the standard FRW
cosmology, such the avoidance of future singularities, are found. At
this point, it should also be noted here that the LQC modification,
although very interesting, is only derived for pure isotropy and is
not very clear how it fits within broader models.

In this paper, we will discuss the dynamics of a phantom scalar
field dark energy, coupled to dark matter in loop quantum cosmology,
to see  in what way the LQG effects would affect the cosmological
evolution of the system. Two kinds of interactions of the form,
$\alpha\rho_m \dot{\phi}$~\cite{case1} and $3\beta
H{(\rho_\phi+\rho_m)}$~\cite{Chimento2003}, will be studied. It is
worth pointing out that the first coupling could arise in string
theory or after a conformal transformation of Brans-Dicke theory,
while the second is motivated by analogy with dissipation of
cosmological fluids and has been proposed for a  possible dynamical
solution to the coincidence problem. Let us note here that the
phantom field, although a viable candidate of dark energy,  also has
some strange properties, such as being unstable to vacuum decay and
the violation of the dominant energy condition.

\section{Loop quantum cosmology}\label{sec2}
By incorporating the effects of loop quantum gravity  which
essentially encode the discrete quantum geometric nature of
spacetime, the effective modified Friedmann equation in a flat
universe is given by~\cite{r44,r53,r54,Sami:2006wj,r60}\footnote{It
is interesting to note that this kind of modified Friedmann equation
also appears in cosmological braneworld models with a single
timelike extra dimension~\cite{Sahni}}
 \be{eq1}
 H^2=\frac{~\rho}{~3}\left(1-\frac{\rho}{\rho_{\rm c}}\right),
 \ee
where $H\equiv\dot{a}/a$ is the Hubble parameter,  $\rho$ is the
total energy density, and  a dot denotes the
 derivative with respect to cosmic time $t$. Here we set $8\pi G\equiv 1$;
 and the critical loop quantum density is
 \be{eq2}
 \rho_c\equiv\frac{\sqrt{3}}{16\pi^2\gamma^3 G^2 \hbar},
 \ee
 where $\gamma$ is the dimensionless Barbero-Immirzi parameter. Let us note here that it
 has been
 suggested that $\gamma\simeq 0.2375$ by the black hole thermodynamics
 in LQG) \cite{Sami:2006wj,Singh:2003au}. Differentiating Eq.~(\ref{eq1}) and using the
  conservation equation of cosmic total energy
 \be{eq3}
 \dot{\rho}+3H\left(\rho+p\right)=0,
 \ee
 one obtains the effective modified Raychaudhuri
 equation
 \be{eq4}
 \dot{H}=-\frac{~1}{~2}\left(\rho+p\right)
 \left(1-2\frac{\rho}{\rho_{\rm c}}\right),
 \ee
 where $p$ is the total pressure. Actually, as shown in~\cite{Magueijo}, the
 effective modified Raychaudhuri equation can be also derived directly by
 using the Hamilton's equations in LQC, without assuming the energy
 conservation.

\section{Dynamics of the interacting phantom scalar field dark energy in LQC}
Let us suppose that there are only the phantom scalar field dark
energy and  dark matter in a spatially-flat universe. The Lagrangian
for the phantom scalar field fluid is
 \begin{equation}
 \mathcal{L} = (1/2)
\partial^\mu \phi
\partial_\mu \phi - V(\phi)\;,
 \end{equation}
where $V(\phi)$ is the potential of the phantom field.  Therefore,
the energy density and pressure for the phantom field can be
expressed as
 \begin{equation}
\rho_\phi=-\frac{1}{2}\dot{\phi}^2+V(\phi)\;,
\end{equation}
 and
\begin{equation}
p_\phi=-\frac{1}{2}\dot{\phi}^2-V(\phi)\;.
 \end{equation}
 We assume that there
is an interaction $\Gamma$
 between the phantom dark energy and
dark matter.  A positive $\Gamma$ corresponds to energy transferring
from phantom to dark matter and vice versa for a negative one.
Therefore the energy densities for the phantom scalar field and dark
matter satisfy the following equations
\begin{eqnarray}
\dot{\rho}_{\phi} + 3H(1+w_{\phi})\rho_{\phi} &=& -\Gamma\,, \label{coupleenerge1} \\
\dot{\rho}_{m} + 3H\rho_{m} &=& +\Gamma\,, \label{coupleenerge2}
\end{eqnarray}
and the Raychaudhuri equation. The evolutionary  equation for the
phantom field and the modified Fiedmann equation can be expressed as
\begin{eqnarray}\label{basic}
\dot{H} &=& -\frac{1}{2} \left(  \rho_m - \dot{\phi}^2
\right)\left(1-{2\,\rho\over \rho_c} \right)\label{7a}\, ,
\end{eqnarray}
\begin{eqnarray}\label{basic2}
\ddot{\phi} &=& -3H\dot{\phi} + V\,' +{\,\Gamma\over {\,\dot
{\phi}}}\,,
\end{eqnarray}
\begin{equation}
H^2  = \frac{1}{3} \left( \rho_m - \frac{1}{2} \dot{\phi}^2 +V
\right)\bigg(1-{\rho \over \rho_c}\bigg) \, , \label{fried}
\end{equation}
 where  $V\,'\equiv dV/d\phi$. In this paper we only consider the case of
exponential potential $V(\phi)=V_0 \exp(-{\lambda } \phi)$ with a
positive constant $\lambda$.

In order to study the dynamics of the above system, we introduce the
following dimensionless variables
\begin{equation}
x\equiv \frac{ \dot{\phi}}{\sqrt{6} H} \, , \quad y \equiv \frac{
\sqrt{V}}{\sqrt{3} H} \,\,\, ,\;\; z\equiv \frac{\, \rho}{\rho_c}
\,,\;\;{d\over dN}\equiv {1\over H}{d\over dt} \label{newvars}\;,
\end{equation}
where $N\equiv\ln{a}$ is the e-folding number and is used as an
independent variable instead of cosmological time.  Using the above
definitions, the effective modified Friedmann equation, namely
Eq.~(\ref{fried}), can be rewritten as
 \be{fried1}
\bigg({\rho_m\over 3H^2}-x^2+y^2\bigg)(1-z)-1=0,
 \ee
and Eq.~(\ref{7a}) becomes
\begin{equation}
{\dot H\over H^2}=-\bigg[{3\over 2}\bigg( {1\over
1-z}+x^2-y^2\bigg)-3x^2\bigg](1-2z) \label{fridmann2}.
\end{equation}
In addition, we will be interested in three scalar quantities, which
are, respectively, the phantom fractional density parameter
$\Omega_\phi$, the effective phantom equation of state $w_\phi$ and
the effective equation of state for total cosmic energy $w_{eff}$
given by
\begin{equation}
\Omega_\phi = -x^2+y^2 \,, \quad w_\phi  = \frac{x^2+y^2}{x^2-y^2},
\,~~~ w_{eff} =(1-z)\bigg[-x^2- y^2\bigg].
\end{equation}

Using the Eqs.~(\ref{basic2}, \ref{fried1}, \ref{fridmann2}), we
obtain the autonomous equations:
\begin{subequations}
\label{autoeqs}
\begin{eqnarray}
x^\prime &=& -3x  - \sqrt{\frac{3}{2}} \lambda\, y^2+{\Gamma\over
\sqrt{6}H^2{\dot{\phi}}} - x \left[3x^2-{3\over 2}\bigg( {1\over
1-z}+x^2-y^2\bigg)\right] (1-2z)\, ,
 \label{autoeqsa} \\
y^\prime &=&  - \sqrt{\frac{3}{2}} \lambda x y -y\bigg[3x^2-{3\over
2}\bigg( {1\over 1-z}+x^2-y^2\bigg)\bigg](1-2z)\, ,
 \label{autoeqsb}\\
z^\prime &=& -3z-3z(1-z)(-x^2-y^2)\,
 .
 \label{autoeqsc}
\end{eqnarray}
\end{subequations}
Let $f\equiv x^\prime, g\equiv y^\prime, h\equiv z^\prime$. Then
using the following condition
\begin{eqnarray}
 (f,~g,~h)|_{(x_c,~y_c,~z_c)}=0~,\label{criticalpoints}
\end{eqnarray}
we can obtain the  critical points $(x_c,~y_c,~z_c)$ of the
autonomous system.

Next, we will study, by examining the stability  of the these
critical points using the standard linearization and stability
analysis, the dynamics of two different interaction cases, i.e.,
$\Gamma_1=\alpha\rho_m {\dot{\phi}}$  and $\Gamma_2=3\beta H (\rho_m
+\rho_\phi)$ between the dark energy and dark matter.

\subsection{Case I: $\Gamma=\alpha\rho_m{\dot{\phi}}$}
This kind of interaction could arise from string theory or
scalar-tensor theory~\cite{case1}. The dynamics of this interacting
phantom model was studied in the standard FRW cosmological framework
in Ref.~\cite{zkguo2} and it was found that energy transfer either
from the phantom field to the dark matter or vice versa yields
similar cosmological consequences, and the energy density of the
phantom field increases with the cosmic expansion, leading to
unwanted future singularity. In this section, we will study the
cosmological evolution of this interacting model in the framework of
loop quantum cosmology. The autonomous Eq.~(\ref{autoeqs}) can be
rewritten as follows
\begin{subequations}
\label{autoeqs2}
\begin{eqnarray}
x^\prime &=& -3x  - \sqrt{\frac{3}{2}} \lambda\, y^2-{\sqrt{6}\over
2}\alpha\bigg( {1\over 1-z}+x^2-y^2\bigg) \nonumber\\&&- x
\left[3x^2-{3\over 2}\bigg( {1\over 1-z}+x^2-y^2\bigg)\right]
(1-2z)\, ,
 \label{autoeqsa2} \\
y^\prime &=&  - \sqrt{\frac{3}{2}} \lambda x y -y\bigg[3x^2-{3\over
2}\bigg( {1\over 1-z}+x^2-y^2\bigg)\bigg](1-2z)\, ,
 \label{autoeqsb5}\\
z^\prime &=& -3z-3z(1-z)(-x^2-y^2)\,\,
 .
 \label{autoeqsc2}
\end{eqnarray}
\end{subequations}
Then we obtain five  critical points:
\begin{eqnarray}
&\bullet&{\rm Point~(A)}:
(\frac{\sqrt{6}\,\alpha}{3},\,~~0\,,~~0\,)\,, \nonumber\\
&\bullet&{\rm Point~(B)}:
\bigg(-\frac{\lambda}{\sqrt{6}},\,~~\sqrt{1+{\lambda^2\over 6}}\,,~~0\,\bigg)\,,  \nonumber\\
&\bullet&{\rm Point~(C)}:
\bigg(-\frac{\lambda}{\sqrt{6}},\,~~-\sqrt{1+{\lambda^2\over 6}}\,,~~0\,\bigg)\,, \\
&\bullet&{\rm Point~(D)}:
\bigg(\frac{\sqrt{6}}{2\,\alpha},\,~~0\,,~~~1-\frac{2\alpha^2}{3}\bigg)\,,  \nonumber\\
&\bullet&{ \rm Point~(E)} :~\bigg(\frac{\sqrt{6}}{2}{1\over
\alpha+\lambda},\,\frac{\sqrt{2\alpha^2-3+2\alpha\lambda}}{\sqrt{2}(\alpha+\lambda)}\,,~0\,\bigg)
\nonumber \,.
\end{eqnarray}
The eigenvalues, $\mu$, of the coefficient matrix of the linearized
equations for these critical points $A, B , C, D$ and $E$ can be
expressed respectively as
\begin{eqnarray}
&&\bullet{\rm ~Point~(A)}:\nonumber\\
&&\mu_1=-\bigg(\alpha^2+{3\over2}\,\bigg),~~~~\mu_2={2\alpha^2-3}
,~~~~\mu_3=-{2\alpha^2-3+2\alpha\lambda\over2},\nonumber \\
&&\bullet{\rm ~Point~(B)}:\nonumber\\
&& \mu_1=\lambda^2,~~~~~~~~\mu_2=-3-{1\over
2}\lambda^2, ~~~~~~~\mu_3=-3-\lambda(\alpha+\lambda),~\nonumber \\
&&\bullet{\rm ~Point~(C)}:\nonumber\\
&& \mu_1=\lambda^2,~~~~~~~\mu_2=-3-{1\over
2}\lambda^2,~~~~~~~ \mu_3=-3-\lambda(\alpha+\lambda),~\nonumber \\
&&\bullet{\rm ~Point~(D)}:\nonumber\\
&& \mu_1=-{3\over2}\bigg(1+\sqrt{1+\frac{6-4
\alpha^2}{\alpha^2}}\bigg),\nonumber\\&&
\mu_2=-{3\over2}\bigg(1-\sqrt{1+\frac{6-4
\alpha^2}{\alpha^2}}\bigg),\nonumber\\&&\mu_3= -{3\over2}{\lambda\over \alpha},\nonumber\\
&&\bullet{\rm Point~(E)}:\nonumber\\
 &&\mu_1={-3\lambda\over
\alpha+\lambda},\nonumber\\&& \mu_2=
-{3\over4}\frac{2\alpha+\lambda}{\alpha+\lambda}\bigg(1+\sqrt{1+\frac{8[3+\lambda(\lambda+\alpha)][2\alpha^2+2\alpha\lambda-3]}
{3[2\alpha+\lambda]^2}}\,\bigg),\nonumber\\
&& \nonumber\\&& \mu_3=
-{3\over4}\frac{2\alpha+\lambda}{\alpha+\lambda}\bigg(1-\sqrt{1+\frac{8[3+\lambda(\lambda+\alpha)][2\alpha^2+2\alpha\lambda-3]}
{3[2\alpha+\lambda]^2}}\bigg)\nonumber.
\end{eqnarray}
In what follows we will analyze the stability of these critical
points:\\
$\bullet$ For point $A$:\\
This fixed point is physically meaningless since $\Omega_\phi=
-{2\alpha^2\over{3}}<0$.\\
$\bullet$ For point $B, C$:\\
The existence of both points are only dependent on $\lambda$. This
can be understood as a result of the fact that the
$\alpha$-dependent term in Eq.~(\ref{autoeqsa}) vanishes when
$\Omega_\phi\rightarrow 1$ and $z\rightarrow 0$. Since the sign of
$\mu_1$ is always opposite to the sign of $\mu_2$, these two points
are saddle points. However, these $\alpha$-independent critical
points are found to be always stable and correspond to the future
singularity in the standard FRW cosmology~\cite{zkguo2}. Therefore
the future
singularity appearing in the standarad cosmology can be avoided by the loop quantum gravity effect. \\
$\bullet$ For point $D$:\\
  This fixed point is physically meaningless since
$\Omega_\phi=-{3\over \alpha^2}<0$.  \\
$\bullet$ For point $E$:\\
The same as in the standard  cosmology, this point is also unstable
in LQC.

From the above analysis, we conclude that there is not any stable
node for this kind of interacting dark energy model in LQC, and the
future singularity appearing in the standard cosmology can be
avoided. Our result of analytical discussions agrees with that of
the numerical calculations obtained in Ref.~\cite{gumjudpai2}.

\subsection{Case II: $\Gamma=3\beta H({\rho_\phi +\rho_m})$}
 This type of interaction is  motivated by analogy with
dissipation of cosmological fluids and has been proposed for a
possible dynamical solution to the coincidence
problem~\cite{Chimento2003}. The dynamics of this interacting
phantom scalar field model in the standard FRW cosmology has been
studied  in  Ref.~\cite{zkguo2} and it was found that there are two
kinds of late time attractors. If the phantom field initially rolls
down the potential the universe will accelerate forever and the
total cosmic energy density decreases with the cosmic expansion,
while if the phantom field initially climbs up its potential the
universe will end with a big rip. Here we discuss the dynamics of
this interacting model in LQC.  For convenience, we introduce
another variable $\xi \equiv{\sqrt{\rho_m}/ (\sqrt{3}H})$. The
autonomous equations  can be rewritten as:
\begin{subequations}
\label{autoeqs3}
\begin{eqnarray}
x^\prime &=& -3x  - \sqrt{\frac{3}{2}} \lambda\, y^2-{3\beta \over
2\,x(1-z)} - x \left[3x^2-{3\over 2}\bigg( {1\over
1-z}+x^2-y^2\bigg)\right] (1-2z)\, ,
 \label{autoeqsa3} \\
y^\prime &=&  - \sqrt{\frac{3}{2}} \lambda x y -y\bigg[3x^2-{3\over
2}\bigg( {1\over 1-z}+x^2-y^2\bigg)\bigg](1-2z)\, ,
 \label{autoeqsb3}\\
z^\prime &=& -3z-3z(1-z)(-x^2-y^2)\,\, ,\\
 \xi^\prime &=&-3\xi\bigg[{1\over
 2}+\bigg(x^2-{1\over2}\,\xi^2\bigg)(1-2z)-\frac{\beta}{2\xi^2(1-z)}\bigg].
 \label{autoeqsc3}
\end{eqnarray}
\end{subequations}
From Eq.~(\ref{fried1}) it is easy to see that there are only three
independent equations in the above system. This system has  four
critical points:
\begin{eqnarray}
Point~A:&&\\\nonumber &&{x_A}^2={1\over2}(\sqrt{1+4\beta}-1),
~~~~~{y_A}=0\,,\\\nonumber &&
{\xi_A}^2={1\over2}(\sqrt{1+4\beta}+1)\,,~~~~~z_A=0\,.\label{solution1}
\end{eqnarray}
Obviously this solution is physically meaningless  since
$\Omega_\phi\,<\,0$. The solutions for the  other three critical
points are tedious and we do not present the details here. However,
these solutions satisfy the following set of equations:
\begin{subequations}
\label{constraints}
\begin{eqnarray}
\beta&=&f(x)\,,\\
y^2&=&-x^2-{\sqrt{6}\lambda x\over3}+1\,,\\
z&=&0\,,\\
\xi^2&=&2x^2+{\sqrt{6}\lambda x\over3}\,,
\end{eqnarray}
\end{subequations}
where we have defined a cubic function
\begin{eqnarray}
f(x)\equiv
x\bigg(2x+{\sqrt{6}\lambda\over3}\bigg)\bigg(1-{\sqrt{6}\lambda
x\over3}\bigg)\,.
\end{eqnarray}
There is a critical point B with $x_B<0$, if
\begin{eqnarray}
0<\beta\leq f\bigg({-\lambda-\sqrt{\lambda^2+12}\over
2\sqrt{6}}\bigg)\,,
\end{eqnarray}
and this point $B$ corresponds to an initially climbing-up phantom
field~\cite{zkguo2}. There are two other critical points ($C$ and
$D$) with $x_{C,D}>0$. One is physically meaningless and we label it
by $D$. Point $C$ exists for
\begin{eqnarray}
0<\beta\leq min\bigg[f\bigg({-\lambda+\sqrt{\lambda^2+6}\over
\sqrt{6}}\bigg)\,,f\bigg({-\lambda+\sqrt{\lambda^2+12}\over
2\sqrt{6}}\bigg)\bigg]\,, \label{RollingDown}
\end{eqnarray}
and it corresponds to an initially rolling-down phantom field.

Employing the standard techniques in the linearization and stability
analysis,  we can obtain three independent evolution equations of
the linear perturbations. Using the eigenvalues of the coefficient
matrix for points $B$ and $C$, found according to
Eq.~(\ref{constraints}), we find that the critical point $B$ is a
saddle point, while  point $C$ is a stable critical point, and it
is, therefore,  a late time attractor.  This is different from what
was obtained in the standard FRW cosmology where both points $B$ and
$C$ are  late time attractors~\cite{zkguo2}. In the standard FRW
cosmology, if the field initially rolls down the potential, the
universe will enter a final state without Big Rip described point
$C$, while for an initially climbing-up phantom field, the universe
will enter a final state described by point $B$ and end with a Big
Rip. Therefore, the future singularity can be avoided by the loop
quantum effect in LQC for the case in which the field initially
climbs up the potential. As a result, in any case, there is no Big
Rip in LQC.   In Fig.~(1) we show the stability regions for
parameter space ($\lambda$, $\beta$). One can see that the effect of
loop quantum gravity breaks the stability of the initially
climbing-up field but leave that of the initially rolling-down field
intact.  This shows that if the universe does not evolve to a  big
rip in the standard cosmology its evolution seems to be uninfluenced
by the loop quantum effect.

Now, we will show results of numerical analysis we have carried out
on this interacting phantom scalar field dark energy model in LQC.
In Figs (\ref{f2}, \ref{f3}) we show the evolution curves of $H$ and
$\rho$ for an initially climbing-up phantom field with different
values of the coupling constant between the dark energy and dark
matter. From these two figures we can see that $H$ climbs up to
reach its maximum value when the total cosmic energy density $\rho$
reaches the value $\frac{\rho_c}{2}=0.75$, and $H$ goes down to zero
when $\rho$ reaches  the maximum $\rho=\rho_c$. After that, the
universe contracts and then bounces. As time goes on, the universe
will undergo oscillations with  increasing frequency which may
eventually blow up. Consequently, this seems to give rise to a new
singularity.  This kind of behavior of possible infinite frequency
of oscillation  also appears in the case of the interaction of the
form $\Gamma=\alpha \dot{\phi}\rho_m$, as found in
Ref.~\cite{gumjudpai2}. In addition, the larger the value of
$\beta$, the later our universe enters the oscillating regime.

In Figs.~(\ref{f4}, \ref{f5}, \ref{f6}, \ref{f7}), we give the
numerical results for an initially rolling-down phantom field with
the requirement given in Eq.~(\ref{RollingDown}). Fig.~(\ref{f4})
shows the evolutionary properties of the universe with  different
initial conditions. Apparently the trajectories converge to the same
final state determined only by parameter $\lambda$. Figs.~(\ref{f5},
\ref{f6}) show the evolutionary curves of $H$ and the effective
equation of state for total cosmic energy $w_{eff}$. We find that
$H$ does not oscillate and  $w_{eff}$ approaches finally to a
constant which is less than $-\frac{1}{3}$ but larger than $-1$.
Thus the universe will keep accelerating forever while its energy
density decreases with the cosmic expansion.  Fig.~(\ref{f7}) shows
that in the final state the energy ratio of the phantom and dark
matter reaches a constant. Therefore the coincidence problem can be
alleviated.

\section{Conclusion}
In conclusion,  we have studied in loop quantum cosmology the
dynamical system of a phantom field coupled to dark matter through
an interaction of the form $\alpha\dot{\phi}\rho_m$ (case I)  or
$3\beta H(\rho_\phi+\rho_m)$ (case II). The exponential potential
for the phantom is used. For case I, there is a late time attractor
solution in the standard FRW cosmology which corresponds to a big
rip of the universe; whereas in LQC this solution transforms to be
unstable, thus the big rip singularity which appears in the standard
cosmology can be avoided by  the loop quantum effect.

For case II,  it was found, in the standard FRW cosmology,  that if
the phantom field is initially rolling down the potential, the
dynamical system has a late time attractor and the universe will
accelerate forever, while if the field initially climbs up  the
potential, there is also a late time attractor but the universe will
end with a big rip. By studying the dynamics of this interacting
phantom model in LQC, we find that the universe with an initially
climbing-up phantom field will oscillate forever; therefore the
future singularity can be avoided by loop quantum effect; while for
an initially rolling-down phantom field the universe will have the
same late time evolution as that in the standard cosmology, i.e.,
the universe will accelerate forever with a constant ratio between
the energies of the phantom and dark matter and the total energy
density of the universe will decrease with the cosmic expansion.
Therefore, in case II,  the loop quantum effect only intervenes when
the universe will evolve to a future singularity, and it sits idle
when otherwise.

\section*{ACKNOWLEDGMENTS}
This work was supported in part by the National Natural Science
Foundation of China under Grants No.10575035, 10775050, the SRFDP
under Grant No. 20070542002, and the Programme for the Key
Discipline in Hunan Province. P. Wu  is partially supported by the
National Natural Science Foundation of China  under Grant No.
10705055,  the Youth Scientific Research Fund of Hunan Provincial
Education Department under Grant No. 07B085, and the Hunan
Provincial Natural Science Foundation of China under Grant No.
08JJ4001.

\begin{figure}[t]
\begin{center}
\includegraphics[width=9cm,height=7.7cm,angle=0]{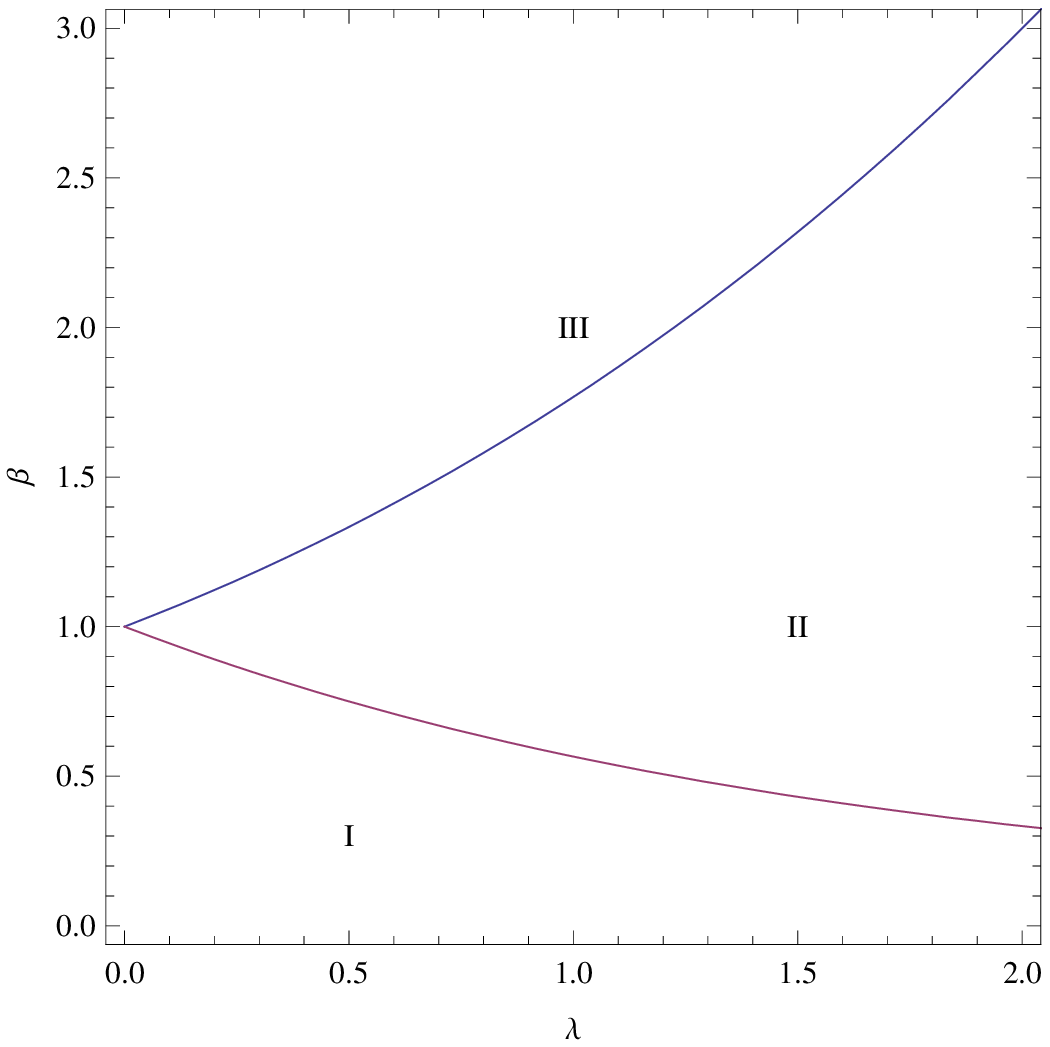}
\end{center}
\caption{The stability regions of ($\lambda$, $\beta$) parameter
space for case II. In LQC,  in the region I  the rolling-down
critical point (Point C) is a stable late time attractor; in region
II there are no stable critical points. However, in the standard FRW
cosmology, in the region I, both the climbing-up scaling solution
and the rolling-down scaling solution are the stable late time
attractors, while in the region  II, the climbing-up solution is the
stable late-time attractor. III represents the region of the
solutions without physical meaning.} \label{region}
\end{figure}

\begin{figure}[t]
\begin{center}
\includegraphics[width=9cm,height=7.7cm,angle=0]{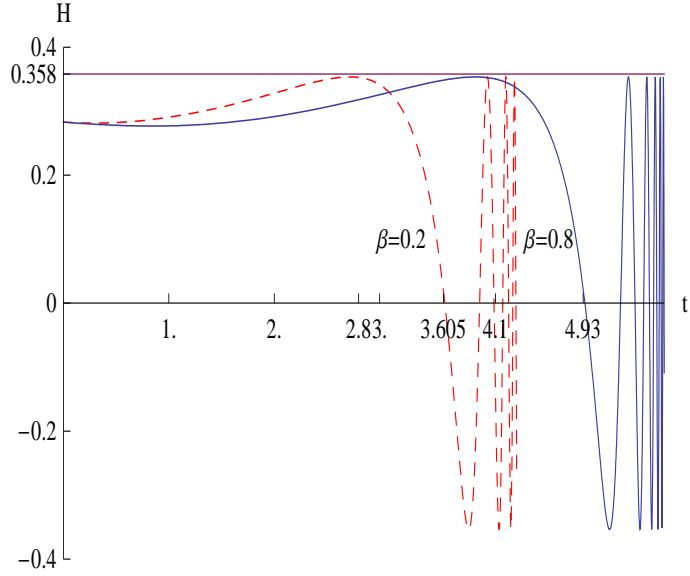}
\end{center}
\caption{The evolution of $H$ for a climbing-up phantom field with
different coupling constants ($\beta=0.2, 0.8$)  and  $\lambda=1,
V_0=1, \rho_c=1.5, \dot{\phi}_0=-0.4$ in case II. } \label{f2}
\end{figure}

\begin{figure}[t]
\begin{center}
\includegraphics[width=9cm,height=7.7cm,angle=0]{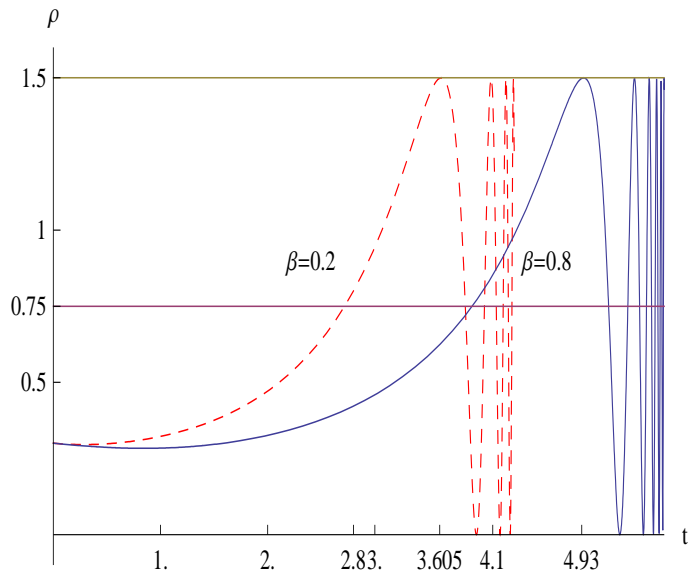}
\end{center}
\caption{The evolution of cosmic total energy density for a
climbing-up phantom field with different coupling constants
$(\beta=0.2, 0.8)$ and  $\lambda=1, V_0=1, \rho_c=1.5,
\dot{\phi}_0=-0.4$ in case II. } \label{f3}
\end{figure}

\begin{figure}[t]
\begin{center}
\includegraphics[width=9cm,height=7.7cm,angle=0]{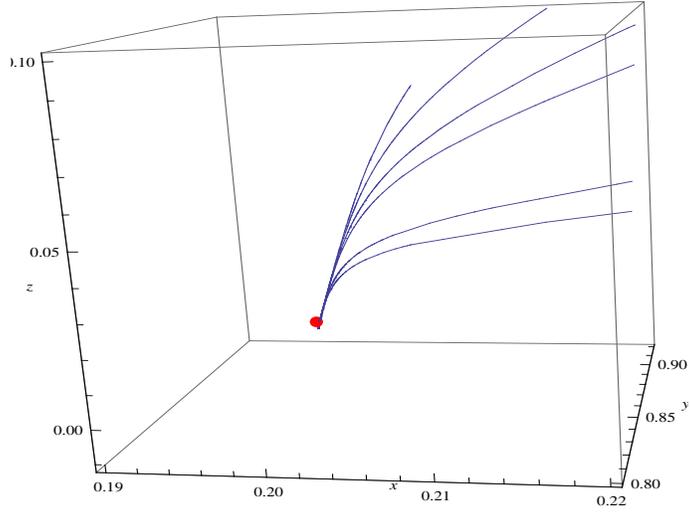}
\end{center}
\caption{The convergence of different initial conditions to the
attractor solution in the ($x,y,z$) phase space for a rolling down
phantom field in case II. The red point denotes the critical point
C} \label{f4}
\end{figure}



\begin{figure}[t]
\begin{center}
\includegraphics[width=9cm,height=7.7cm,angle=0]{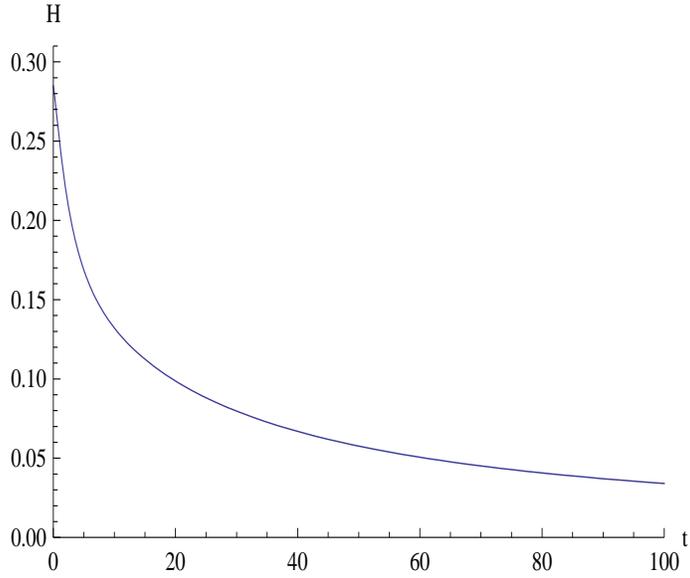}
\end{center}
\caption{The evolution of $H$ with time for a rolling-down phantom
field with $\beta=0.2, \lambda=1, V_0=1, \rho_c=1.5$ and
$\dot{\phi}_0=0.4$. } \label{f5}
\end{figure}


\begin{figure}[t]
\begin{center}
\includegraphics[width=9cm,height=7.7cm,angle=0]{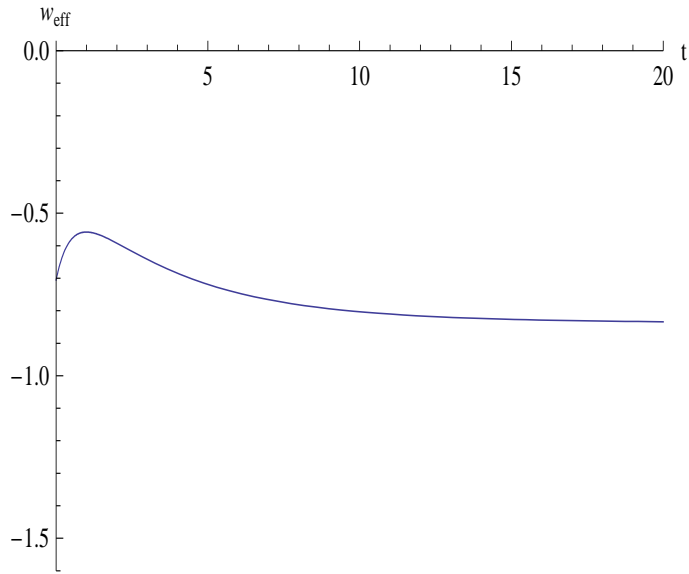}
\end{center}
\caption{The evolution of the effective equation of state for the
total cosmic energy in a rolling-down phantom model with $\beta=0.2,
\lambda=1, V_0=1, \rho_c=1.5$ and $\dot{\phi}_0=0.4$. } \label{f6}
\end{figure}

\begin{figure}[t]
\begin{center}
\includegraphics[width=9cm,height=7.7cm,angle=0]{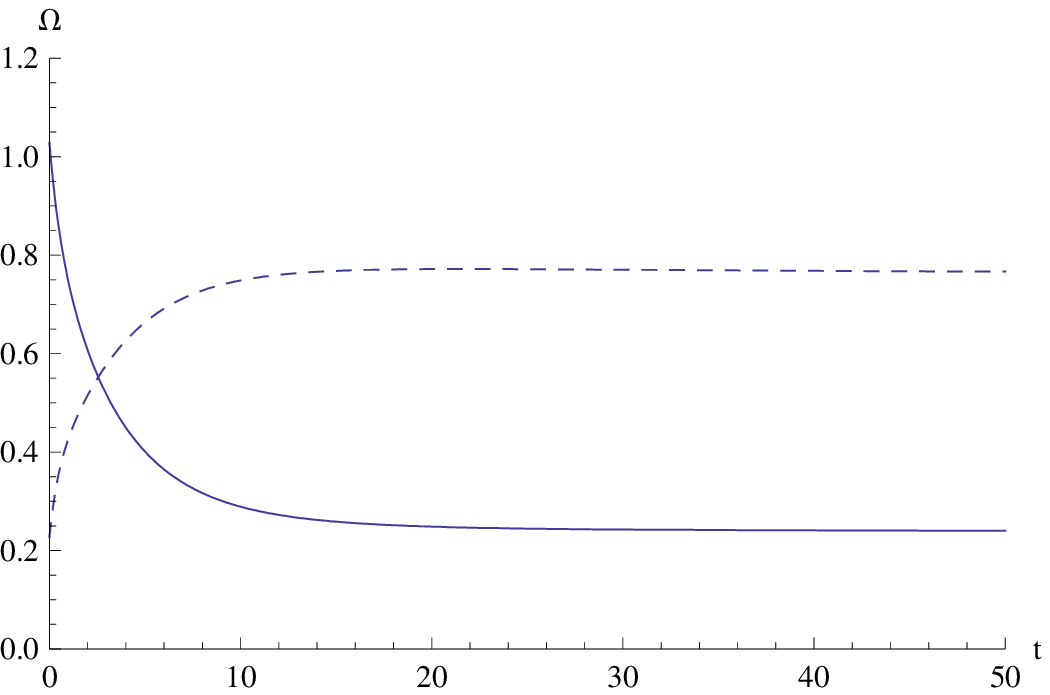}
\end{center}
\caption{The evolutionary curves of the fractional densities of a
phantom field (dashed line) and  dark matter (solid line) for a
rolling-down phantom model with $\beta=0.2, \lambda=1, V_0=1,
\rho_c=1.5$ and $\dot{\phi}_0=0.4$. } \label{f7}
\end{figure}





\begin{thebibliography}{99}


\bibitem{bennett}C. L. Bennett, {\it et al}., Astrophys. J. Suppl. {\bf 148},1 (2003) [arXiv:astro-ph/0302207];\\
                 D. N. Spergel, {\it et al}., Astrophys. J. Suppl. {\bf 148},175 (2003) [arXiv:astro-ph/0302209];\\
                 S. Masi, {\it et al}., Prog. Part. Nucl. Phys. {\bf 48}, 243 (2002) [arXiv:astro-ph/0201137].

\bibitem{scranton} R. Scranton,  {\it et al}.,  [arXiv:astro-ph/0307335];\\
                 A. G. Riess, {\it et al}., Astron. J.  {\bf 116}, 1009 (1998) [arXiv:astro-ph/9805201];\\
                 S.  Perlmutter, {\it et al}., Astrophys. J. {\bf 517}, 565 (1999)  [arXiv:astro-ph/9812133];\\
                 G.  Goldhaber, {\it et al}.,  [arXiv:astro-ph/0104382].

\bibitem{weinberg}S. Weinberg, Rev. Mod. Phys. {\bf 61}, 1 (1989);\\
                  V. Sahni and A. Starobinsky, Int. J. Mod. Phys. D {\bf 9}, 373 (2000); \\
                  P. J. E. Peebles and B. Ratra, Rev. Mod. Phys. {\bf 75}, 559 (2003);\\
                  T.  Padmanabhan, Phys. Rept. {\bf 380}, 235
                  (2003).\\
                  E.~J.~Copeland, M.~Sami and S.~Tsujikawa,  Int.\ J.\ Mod.\ Phys.\  D {\bf 15}, 1753
                  (2006).


\bibitem{Quint}   C. Wetterich, Nucl. Phys. B {\bf 302},  668 (1988);
                  B. Ratra and  P. E. J. Peebles, Phys. Rev.  D  {\bf 37},  3406 (1988);
                  R. Caldwell,  R. Dave, and  P. J.  Steinhardt, Phys. Rev. Lett. {\bf 80},  1582 (1998).

\bibitem{Cald}    R. R. Caldwell,  Phys. Lett. B {\bf 545}, 23 (2002);
                  R. R. Caldwell, M. Kamionkowski, and  N. N. Weinberg,  Phys. Rev. Lett. {\bf 91}, 071301 (2003);
                  P. Singh et al, Phys. Rev. D {\bf 68},023522 (2003);
                  S. Nesseris and  L. Perivolaropoulos,  Phys. Rev. D {\bf 70},  123529  (2004).
                  S. Nojiri and S. D.  Odintsov,  Phys. Lett. B {\bf 571}, 1 (2003);
                  S. Nojiri and S. D.  Odintsov,   Phys. Lett. B {\bf 562}, 147  (2003);
                  S. Nojiri and S. D.  Odintsov,  Phys.  Rev. D {\bf 70}, 103522  (2004);
                  S.  Nojiri, S. D. Odintsov, and  S. Tsujikawa,   Phys. Rev. D {\bf 71}, 063004  (2005);
                  P.  Wu and H. Yu, J. Cosmol. Astropart. Phys. {\bf 05}, 008 (2006);
                  S.M. Carroll, M. Hoffman, M. Trodden, Phys. Rev. D {\bf 68}, 023509 (2003);
              J. Cline, S. Jeon, G. Moore, Phys. Rev. D {\bf 70},  043543 (2004);
              B. McInnes,   J. High  Energy  Phys. {\bf 0208},  029 (2002);
              V. Sahni, Y. Shtanov, J. Cosmol. Astropart. Phys. {\bf 0311},  014 (2003);
              P.F. Gonzalez-Diaz, Phys. Rev. D {\bf 68},  021303 (2003);
              M. Bouhmadi-Lopez, J.A. Jimenez Madrid, J. Cosmol. Astropart. Phys. {\bf 0505}, 005 (2005). 
              E. Elizalde, S. Nojiri, S.D. Odintsov, Phys. Rev. D {\bf 70},  043539 (2004).
              P.  Wu and H. Yu,    Nucl. Phys.  B {\bf 727}, 355
              (2005);
              M.R. Setare,  Eur. Phys. J. C {\bf 50}, 991 (2007);
              B. Boisseau, G. Esposito-Farese, D. Polarski,  A. Starobinsky, Phys. Rev. Lett. {\bf 85}, 2236
              (2000).

\bibitem{Quintom}B. Feng, X. Wang and X. Zhang, Phys. Lett. B {\bf 607}, 35 (2005);
                 B. Feng, M. Li, Y. Piao and X. Zhang, Phys. Lett. B {\bf 634},  101 (2006); 
                 Z. Guo, Y. Piao, X. Zhang and Y. Zhang, Phys. Lett. B {\bf 608}, 177 (2005);
                 P. Wu and H. Yu, Int. J. Mod. Phy. D {\bf 14}, 1873 (2005);
                 Z. Guo, Y. Piao, X. Zhang and Y. Zhang, Phys. Rev. D {\bf 74}, 127304 (2006);  
                 R. Lazkoz, G. Leon, I. Quiros, Phys. Lett. B {\bf 649}, 103 (2007); 
                 R. Lazkoz, G. Leon,  Phys. Lett. B {\bf 638},   303 (2006); 
                 Y. Cai, T. Qiu, Y. Piao, M. Li and  X. Zhang, J. High  Energy  Phys. {\bf 0710}, 071 (2007); 
                 Y. Cai, M. Li, J. Lu, Y. Piao, T. Qiu and X. Zhang,  arXiv:hep-th/0701016[astro-ph];
                 Y. Cai, H. Li, Y. Piao and X. Zhang,  Phys. Lett. B {\bf 646}, 141 (2007);
                 G. Zhao, J. Xia, M. Li, B. Feng and X. Zhang, Phys.Rev. D {\bf 72},  123515  (2005);
                  M. R. Setare, Phys. Lett. B {\bf 641}, 130 (2006);
                 X. Zhang and T. Qiu,  Phys.Lett. B {\bf 642},  187 (2006).
\bibitem{Hessence}H. Wei, R. G. Cai and D. F. Zeng, Class. Quant. Grav. {\bf 22}, 3189 (2005);
                   H. Wei and R. G. Cai, Phys. Rev. D {\bf 72}, 123507 (2005);
                   M. Alimohammadi and H. Mohseni Sadjadi, Phys. Rev. D {\bf 73}, 083527 (2006);
                   W. Zhao and Y. Zhang, Phys. Rev. D {\bf 73}, 123509 (2006);
                   H. Wei, N. N. Tang and S. N. Zhang, Phys. Rev. D {\bf 75}, 043009
                   (2007).


\bibitem{Chimento2003}L. P. Chimento, A. S. Jakubi, D. Pavon and W. Zimdahl,  Phys. Rev. D {\bf 67},  083513 (2003);
                      L. P. Chimento and  D. Pavon, Phys. Rev. D {\bf 73}, 063511 (2006).



\bibitem{RovelliThiemann}C. Rovelli, Living Rev. Rel. {\bf 1}, 1 (1998) [arXiv:gr-qc/9710008];\\
                         T. Thiemann, Lect. Notes Phys. {\bf 631}, 41 (2003) [arXiv:gr-qc/0210094];\\
                         A. Corichi, J. Phys. Conf. Ser. {\bf 24}, 1 (2005) [arXiv:gr-qc/0507038];\\
                         A. Perez, arXiv:gr-qc/0409061.

\bibitem{Ashtekar2}A. Ashtekar and J. Lewandowski, Class. Quant. Grav. {\bf 21}, R53 (2004) [arXiv:gr-qc/0404018];\\
                  A. Ashtekar, arXiv: 0705.2222[gr-qc].

\bibitem{Rovelli} C. Rovelli, {\it Quantum Gravity}, Cambridge University Press, Cambridge (2004).

\bibitem{Ashtekar3} A. Ashtekar, New J. Phys. {\bf 7}, 198 (2005) [arXiv:gr-qc/0410054];\\
                    T. Thiemann, hep-th/0608210.

\bibitem{r42}M. Bojowald, Living Rev. Rel.  {\bf 8}, 11 (2005) [arXiv:gr-qc/0601085];\\
            M. Bojowald,  arXiv:gr-qc/0505057.
\bibitem{Mielczarek} J. Mielczarek, T. Stachowiak, M. Szydlowski, arXiv: 0801.0502v2
[gr-qc].
\bibitem{r43}A. Ashtekar, M. Bojowald and J. Lewandowski, Adv. Theor. Math. Phys. {\bf 7}, 233 (2003) [arXiv:gr-qc/0304074];\\
            A. Ashtekar,  arXiv:gr-qc/0702030.

\bibitem{r44}A. Ashtekar, AIP Conf. Proc. {\bf 861}, 3 (2006) [arXiv:gr-qc/0605011].


\bibitem{r54}P. Singh, Phys. Rev.  D {\bf 73}, 063508 (2006) [arXiv:gr-qc/0603043].


\bibitem{r53} A. Ashtekar, T. Pawlowski and P. Singh,  Phys. Rev.  D {\bf 74}, 084003 (2006) [arXiv:gr-qc/0607039].

\bibitem{Bojowald08} M. Bojowald, arXiv: 0801.4001[gr-qc].

\bibitem{SA} A. Corichi and P. Singh, Phys. Rev. Lett {\bf 100}, 161302
(2008)); A. Ashtekar et al, Phys. Rev.  D {\bf 75}, 024035 (2007).


\bibitem{Sami:2006wj}M. Sami, P. Singh and S. Tsujikawa,  Phys. Rev.  D {\bf 74}, 043514 (2006) [arXiv:gr-qc/0605113];
                     T. Naskar and J. Ward, arXiv:0704.3606 [gr-qc].

\bibitem{Singh:2003au}P. Singh and A. Toporensky, Phys. Rev. D {\bf 69}, 104008 (2004)
[arXiv:gr-qc/0312110].

\bibitem{Daris} D. Samart and Burin Gumjudpai,  Phys. Rev. D {\bf 76}, 043514 (2007) [arXiv:hep-th/0704.3414].

\bibitem{Hao} H. Wei and S. N. Zhang,  Phys. Rev. D {\bf 76}, 063005 (2007)
[arXiv:gr-qc/0705.4002].

\bibitem{WuZhang}P. Wu and S. N. Zhang, [arXiv:astro-ph/0805.2255].

\bibitem{case1} C. Wetterich, Astron. Astrophys. {\bf  301}, 321 (1995);\\
                L. Amendola, Phys. Rev. D {\bf 60}, 043501  (1999).


\bibitem{r60}X. Zhang and Y. Ling, arXiv:gr-qc/0705.2656.

\bibitem{Sahni} Y. Shtanov and V. Sahni,  Phys. Lett. B {\bf557}, 1 (2003).


\bibitem{Magueijo}J. Magueijo and P. Singh,  arXiv:astro-ph/0703566.


\bibitem{zkguo2}Z. K. Guo, R. G. Cai and  Y. Z. Zhang, JCAP {\bf 05}, 002 (2005) [arXiv:astro-ph/0412624v2].



\bibitem{gumjudpai2}B. Gumjudpai,  arXiv:gr-qc/0706.3467v2.




\end{thebibliography}
\end{document}